\documentclass[sigconf]{acmart}

\AtBeginDocument{%
  }

\acmConference[ICMHI '26]{10th International Conference on Medical and Health Informatics}{May 15--17,
  2026}{Kyoto, Japan}

\begin{document}

\title{Modeling and Interpreting Teamwork Dynamics in Cancer Care Outcome Prediction}

\author{Yuhua Huang}
\authornote{Corresponding Author}
\email{yuuhuang@ucdavis.edu}
\affiliation{%
  \institution{University of California, Davis}
  \city{Davis}
  \state{California}
  \country{USA}
}
\author{Hsiao-Ying Lu}
\email{hyllu@ucdavis.edu}
\affiliation{%
  \institution{University of California, Davis}
  \city{Davis}
  \state{California}
  \country{USA}
}
\author{Kwan-Liu Ma}
\email{klma@ucdavis.edu}
\affiliation{%
  \institution{University of California, Davis}
  \city{Davis}
  \state{California}
  \country{USA}
}

\renewcommand{\shortauthors}{Huang et al.}

\begin{abstract}
    Cancer care requires a longitudinal approach in which treatments are planned and delivered over time according to the needs of each individual patient. While prior research has thoroughly explored how clinical and demographic factors—such as comorbidities and age—inform treatment planning, far less attention has been devoted to the delivery phase of care. Yet planning and delivery are both team-based processes that depend on coordinated efforts among multiple healthcare professionals (HCPs). As such, the human factors embedded in these collaborative practices are crucial to optimizing patient outcomes. Despite this importance, the existing literature on human factors in cancer care is limited, and very few studies have investigated how collaboration within care teams evolves over the course of treatment.
    To fill this gap, this work examine how HCPs' collaboration—captured through electronic health record (EHR) systems—affects cancer patient outcomes, with particular emphasis on teamwork dynamics. We represent EHR-mediated HCP interactions as networks and apply machine learning methods to identify predictive signals of patient survival embedded in these collaborative structures. We further interpret model predictions by pinpointing network characteristics and dynamic patterns associated with particular outcomes.
    We evaluate our model through robustness analyses to ensure that the findings are stable and not driven by stochastic variation in training. Additionally, our insights align with hypotheses proposed in the medical literature, and our results provide the empirical, data-driven evidence supporting these claims. Overall, our work contributes a practical workflow for leveraging digital traces of collaboration to evaluate and strengthen longitudinal team-based healthcare. The approach is potentially transferable to other domains involving complex collaboration and offers actionable insights to guide data-informed interventions in healthcare delivery.
\end{abstract}


\maketitle

\section{Introduction}
Cancer care unfolds through a series of complex clinical decisions and coordinated actions that span an extended treatment trajectory. While patient characteristics such as age, disease stage, and comorbidities have long been recognized as important predictors of cancer outcomes \cite{sogaard2013impact, brandt2015age}, they represent only part of the picture. Cancer treatment rarely occurs in isolation; instead, it involves multiple healthcare professionals (HCPs) working together to plan and deliver care. Research in human factors has shown that team coordination plays a vital role in care quality in complex medical settings \cite{gurses2006systematic, smits2010exploring, bagnasco2013identifying, verhaegh2017exploratory}, suggesting that collaboration among HCPs may also influence clinical outcomes. However, this dimension remains largely underrepresented in predictive modeling for oncology.

Only a small portion of existing studies directly investigate how teamwork quality may relate to cancer outcomes. Prior work has taken initial steps in this direction. For example, \cite{lu2025associating} demonstrated that there are associations between collaboration mediated by the electronic health system (EHR) and patient survival, along with identifying structural traits linked to favorable outcomes. A separate effort \cite{lu2024ehrflow} examined HCP collaboration patterns using interpretable network measures and showed how these patterns vary throughout treatments. Although these studies highlight the relevance of teamwork, they do not specify how changes in collaboration over time relate to better or worse survival outcomes, leaving the temporal dimension of team dynamics insufficiently characterized.

To address this gap, we model HCP interactions as temporal collaboration networks derived from time-stamped EHR activity. We introduce a data-driven framework that leverages machine learning to capture predictive signals embedded in evolving collaboration structures, and we interpret these signals through network dynamics that unfold along the treatment timeline. This temporal perspective enables us not only to identify which collaboration features are associated with patient survival, but also when those features become influential. We evaluate this approach with robustness analyses to confirm that the findings are stable and not driven by training randomness. Our results also support theories proposed in the clinical literature and offer the empirical, large-scale validation of these theories using EHR collaboration traces.

In summary, our contributions are twofold: (1) we propose a transferable framework for interpreting dynamic predictive signals of cancer survival in temporal networks of EHR-mediated collaboration, and (2) we demonstrate that the learned features provide actionable insights for improving care delivery.
More broadly, this work illustrates how digital traces of teamwork, when combined with machine learning, can support the assessment and optimization of team-based cancer care, with potential applicability to other collaborative healthcare domains.
\section{Related Works}
Research on predicting cancer outcomes has predominantly centered on individual patient characteristics—including demographic factors, comorbidities, cancer stage, and prior treatments \cite{piccirillo2004prognostic, sogaard2013impact, brandt2015age, nixon1994relationship}. Although these predictors have enabled useful survival models, they tend to conceptualize patients independently of the care environment, paying little attention to how treatment is coordinated and delivered by teams of clinicians.

At the same time, a parallel line of work highlights the importance of human influences in shaping patient outcomes. Studies in this space have examined healthcare professionals' communication structures, role distributions, and patterns of teamwork within clinical settings \cite{gurses2006systematic, smits2010exploring, bagnasco2013identifying, verhaegh2017exploratory}. Yet, despite these efforts, the contribution of collaborative processes to clinical performance remains underinvestigated.

With the widespread adoption of electronic health records (EHRs), new opportunities have emerged to observe and measure collaboration directly. EHR systems now serve as the backbone of documentation and care coordination, producing detailed, time-stamped records of clinical activities. While prior research has largely leveraged EHR data to analyze patient-level physiological or diagnostic information \cite{huang2017regularized, amirahmadi2023deep, nelson2019integrating}, the metadata created through system interactions has been underutilized. Digital footprints—such as record access logs, shared notes, and co-signature behaviors—offer a rich view of how care teams operate in practice, revealing coordination behaviors at scale without relying on manual observation or survey-based reporting.

Machine learning methods have been leveraged in healthcare studies, powering applications such as forecasting disease trajectories, personalizing treatments, and identifying high-risk patients \cite{uddin2019comparing, atan2018deep, ballinger2018deepheart, rath2022prediction}. Recent work has also advanced explainability techniques to make these models more transparent for clinical adoption \cite{mienye2024optimized, alsaleh2023prediction}. Nonetheless, the vast majority of studies continue to focus on signals tied directly to patients rather than the human and organizational factors that shape care planning and delivery. 

In contrast, \cite{lu2025associating} investigated the use of machine learning to model EHR-mediated collaboration among HCPs, addressing methodological barriers related to capturing digital teamwork and extracting interpretable quality signals from it. However, this work does not consider the dynamics of collaboration, in which HCPs and clinical documentations joined dynamically throughout the treatment timeline. Another study \cite{lu2024ehrflow} approached teamwork dynamics through visual analytics, enabling interactive exploration of how collaboration structures change over time. While informative, this approach facilitates exploration rather than directly analyzing how temporal collaboration patterns influence patient outcomes.
To fill this gap, our work directly models and analyzes the longitudinal course of cancer care, interpreting dynamic collaboration patterns that are associated with patient survival.
\section{Background}
\label{sec:bkgd}
We leverage digital traces embedded in the EHR system to construct evolving collaboration networks among HCPs for each individual patient. These temporal networks capture how care teams form and interact over time, reflecting the dynamic nature of cancer treatment delivery. We then employ a temporal graph neural network (TGNN) to extract predictive signals of patient survival from these evolving collaboration structures. In this section, we provide a brief background on temporal graphs and TGNNs.

    \paragraph{Temporal Graphs}
    A temporal graph is a time-evolving graph defined as $\mathcal{G} = \{G_1, G_2, \ldots, G_T\}$, where each snapshot $G_t = (V_t, E_t)$ represents the system state at time $t$. 
    Nodes $V_t$ correspond to entities and edges $E_t$ encode interactions that occur within the given time window. 
    % In our context, each patient is associated with a sequence of collaboration graphs in which nodes represent HCPs and , and edges denote collaborative activities captured through time-stamped EHR interactions (e.g., shared documentation, co-signatures, sequential access). 
    % This formulation captures how care teams form and evolve during the longitudinal course of treatment.

    \paragraph{Temporal Graph Neural Networks (TGNNs)}
    Graph neural networks (GNNs) learn node or graph-level representations by aggregating information from neighbors through message passing. 
    Temporal graph neural networks extend this paradigm by modeling the dynamic nature of edges and node features across time. 
    A TGNN learns a representation $h_t$ for each snapshot $G_t$ and integrates these representations to capture temporal dependencies:
    \begin{equation}
        h_t = \mathrm{GNN}(G_t), \qquad
        H = \mathrm{TemporalAgg}(h_1, h_2, \ldots, h_T),
    \end{equation}
    where $\mathrm{TemporalAgg}(\cdot)$ may consist of recurrent networks (e.g., GRU, LSTM), attention mechanisms, or time-aware convolutional modules. 
    The resulting representation $H$ encodes both structural collaboration patterns and their temporal evolution, enabling downstream tasks such as classification or prediction.

\section{Methodology}
\label{sec:method}

    \subsection{Data: EHR-Mediated HCP Collaborations}
    Building upon the data cleaning and processing procedures proposed in \cite{lu2025associating}, we additionally model the temporal evolution of the collaboration data.
        \subsubsection{Data overview.}
        \label{sec:data}
            Our raw data consists of EHR digital traces from 505 patients diagnosed with Stage 2 or 3 breast, lung, and colorectal cancers, with approval from the IRB for data use. For each patient, the dataset includes their \textit{basic information} and \textit{access logs} of their EHR data. Basic information encompasses demographics (e.g., age and gender), treatments, comorbidities, and survival outcome (alive/dead). EHR access logs contain timestamped events spanning three months before to one year after the diagnosis date. To ensure a consistent collaboration timeframe, we exclude patients who passed away within a year of diagnosis.
            % To maintain a consistent collaboration timeframe, we exclude patients who passed away within a year after diagnosis.
            A timestamped EHR access event involves a HCP accessing (reviewing or writing) a document, such as a note or a message, on the EHR system. Since HCPs typically record a patient's medical conditions in notes, while messages often lack context, we include only the access events involving notes in our analysis. 
            % Therefore, throughout this paper, we use the terms ``note'' and ``HCP/EHR document'' interchangeably. 
            To focus on assessing the interactions within core teams, following the recommendations of our medical doctor collaborators, we include only HCPs with the titles MD, NP, PA, RN, Pharmacy Technician, Pharmacist, and Case Manager.

            % This data captures the flow of information among HCPs, allowing us to track which HCP authored a note and who subsequently read it. Those who read the note may then write additional notes, further disseminating the acquired information. We extract these collaborative interactions that enable information transfer. To ensure a consistent collaboration timeframe, we exclude patients who passed away within a year of diagnosis.
    
        \subsubsection{Data processing and categorization.}
        \label{sec:featureVar}
            To extract the collaboration surrounding a patient, we identify all notes related to the patient and the HCPs who have reviewed, written, or edited these notes using EHR access logs. Each note is further characterized by three variables: the category of the note's intent, the category of its content, and a label indicating whether it was created during an inpatient period. There are five intent categories, including \textit{Orders} and \textit{Patient Clinical Information}, and 32 content categories, such as \textit{Order Canceled} and \textit{Note Signed}.
            For each HCP, we provide context into their role in the collaboration using four variables: title, type, specialty, and a label indicating whether they are a resident. There are seven titles (e.g., \textit{MD} or \textit{RN}), 12 types (e.g., \textit{Physician Faculty} or \textit{Physician Fellow}), and 71 specialties (e.g., \textit{Cardiology} or \textit{Dermatology}).
            
        \subsubsection{Bipartite network construction.}
            After identifying all participants in the collaboration surrounding a patient (notes and HCPs), we define the information flow among these entities. This flow is represented as a directed bipartite network, where notes and HCPs serve as network nodes, and edges capture the reviewing and writing events recorded in the EHR system. For example, if $HCP_A$ reviews $Note_B$, an edge is established from $Note_B$ to $HCP_A$, indicating that the information from $Note_B$ flows toward $HCP_A$.
            The bipartite nature of this collaboration network ensures that edges only form between a note and an HCP, but not between two notes or two HCPs. This reflects the fact that interactions among HCPs within the EHR are always mediated through notes rather than direct communication. Finally, the variables extracted to characterize each note and HCP, as described in \autoref{sec:featureVar}, are assigned as node attributes. 
            % The constructed collaboration networks are illustrated in \figref{fig:networkExample}.

        \subsubsection{Network snapshot sequence construction.}
        \label{sec:snapshot}
            To capture how collaboration evolves over time, we transform each patient's time-stamped interactions into a sequence of graph snapshots.
            For every note--HCP edge (an edge between a note node and an HCP node), we collect the timestamps at which the interaction occurred in the access logs.
            Using the earliest interaction time $t_0$ for a given patient as a reference, we partition the timeline into fixed-length, non-overlapping one-week intervals of duration $\Delta t$ and assign each interaction at time $t$ to a bin $b = \lfloor (t - t_0) / \Delta t \rfloor$.
            
            From these binned events, we construct a sequence of \emph{cumulative} bipartite graphs.
            We iterate over the bins in chronological order and, for each bin $b_t$, create a network snapshot $G_t$ that includes all note--HCP edges that have appeared in any bin up to and including $b_t$ (i.e., all edges in $b_{0}, \dots, b_{t}$).
            % edited  
            Consequently, the active care team grows over time: in early snapshots only a few HCPs and notes participate, and in later snapshots additional nodes become active as they first appear in the access logs. The edge set is cumulative, so once a HCP has appeared, their past interactions are retained in all subsequent snapshots.
            % PREV: Consequently, the node set for a patient remains fixed across time, while the edge set monotonically grows as the care team's collaboration expands.
           
            %===================
            % edited 
            For each snapshot $G_t$, we construct a node feature matrix $X_t \in \mathbb{R}^{N_t \times F}$, where $N_t$ is the number of nodes and $F$ is the feature dimension.
            Each node $v$ has a feature vector
            \[
            x_v^{(t)} = [1,\ \deg_t(v),\ a_v],
            \]
            where the first entry is a fixed bias term (set to $1.0$ for all nodes), the second entry is the degree of node $v$ in $G_t$, and $a_v \in \mathbb{R}^{D_{\text{attr}}}$ is the attribute vector of node $v$, as detailed in \autoref{sec:featureVar}.
            In this case, the feature dimension is $F = 2 + D_{\text{attr}}$, which remains fixed across snapshots, while the number of nodes $N_t$ may grow over time as more HCPs and notes become active.
            These enriched node features serve as the input to the temporal graph neural network, and thus the per-node attribute matrix $a_v$ is directly used as initial feature by the predictive model.
            
            % PREV : For each snapshot, we compute structural node features consisting of a constant term and the node’s degree within that snapshot; these two-dimensional features serve as the input to the temporal graph neural network.
            % In parallel, we attach a per-node attribute matrix (provider type, title, residency status, specialty, note metric description, metric group, and inpatient flag) derived from the raw logs.
            %===================
            %edited 
           % The per-node attribute matrix $a_v$ thus directly contributes to the initial node features used by the predictive model.
            %PREV:  This matrix is utilized for interpretability analysis but is not directly fed into the predictive model.

    \subsection{Survival Prediction using Temporal Graph Neural Network}
    \label{sec:predictMethod}
        % predictive target variable, time window
        % evolveGCN architecture
        We formulate outcome prediction as a patient-level binary classification task: given a patient’s temporal collaboration network, the model predicts whether the patient survives at the end of the analysis timeframe described in \autoref{sec:data}. To account for intrinsic collaboration differences driven by clinical concerns across diseases, we train separate models for breast, lung, and colorectal cancer cohorts.
    
        We employ EvolveGCN \cite{pareja2020evolvegcn}, a temporal extension of graph convolutional networks designed for evolving graphs. As discussed in \autoref{sec:bkgd}, EvolveGCN takes as input a sequence of bipartite network snapshots for each patient. At each time step $t$, a graph convolutional network (GCN) processes the current snapshot $G_t$ to learn node embeddings that capture the local collaboration context. EvolveGCN then updates the internal parameters of the GCN through a recurrent mechanism, allowing the GCN to adapt as the collaboration structure changes over time. 
        To be specific, we adopt the EvolveGCN-O variant, where a gated recurrent unit (GRU) is used to evolve the weight matrices of each GCN layer over time.
        % In the EvolveGCN-O variant used here, a gated recurrent unit (GRU) evolves the weight matrices of each GCN layer over time, so that the convolution filters themselves become a function of the collaboration history
        % PREV: EvolveGCN then updates the internal parameters of the GCN through a recurrent mechanism, allowing GCN to adapt as the collaboration structure changes over time. In our experiments, we use the EvolveGCN-O variant.
    
        To construct a network-level representation, the node embeddings within each snapshot are aggregated using a permutation-invariant readout (global mean pooling), producing a single vector summarizing collaboration traits at that time step. These snapshot-level embeddings are then further aggregated across time either by mean pooling or by taking the final snapshot embedding. The resulting network-level representation is passed to a lightweight feed-forward classifier that outputs the predicted survival probability.
        
        The EvolveGCN model is trained using a binary cross-entropy loss associated with each patient’s survival label. Class imbalance (i.e., skewed alive and deceased patient distribution) within each cancer cohort is addressed by oversampling the minority class. We convert predicted probabilities into binary labels using a decision threshold selected via a constrained grid search that maximizes validation accuracy. Final performance is assessed using F1 score, accuracy, precision ,and recall on a held-out test set. 
        % We also report five-fold cross-validation results to evaluate generalizability to unseen collaboration networks.

    \subsection{Interpreting Temporal Graph Neural Network}
    \label{sec:interpretMethod}
        % attribute surrogate explanation method / Shapley method (pending)
        % use the attribute explanations to find the temporal segmentation point
        % pre transition post performance comparison (expectation: increasing if positively correlated, decreasing otherwise)
        % while evolvegcn is more effective due to its evolving technique, it is more opaque. We cannot use the usual attention score as a proxy of the snapshot importance...

        While TGNNs offer strong predictive performance, they often function as black boxes because their internal nonlinear transformations are not directly interpretable. In the case of EvolveGCN, its evolving mechanism—where recurrent updates adjust the GCN parameters over time—can further increase opacity by making it difficult to trace how learned representations change across collaboration snapshots. As a result, connecting model decisions back to clinically meaningful teamwork patterns is nontrivial.
        To address this challenge, we employ an attribute-based surrogate model together with a temporal segmentation procedure. This combination enables us to identify which collaboration characteristics are influential and when they exert their effects along the treatment timeline.

        \paragraph{Attribute-based surrogate.}
        We begin by summarizing the node attributes defined in \autoref{sec:featureVar}. For HCP nodes, we compute the frequency of each attribute—for instance, the number of times a \emph{Cardiologist} appears in the temporal network divided by the total number of HCP nodes—providing a measure of how strongly each role participates in the patient’s care. For note nodes, we apply the same frequency calculation to attributes such as \emph{Orders} or \emph{Patient Clinical Information}, capturing the extent to which different types of documentation are involved.
        For each cancer cohort, we train a simple linear classifier (logistic regression) using these attribute summaries as a surrogate for the trained EvolveGCN. The surrogate performs the same prediction task but operates in a semantically meaningful and interpretable feature space. Because of its linear structure, the learned weight assigned to each attribute summary directly indicates its contribution to the patient outcome: positive weights suggest factors associated with better survival, whereas negative weights highlight factors that may require intervention. This enables us to interpret which HCP roles and note types are most strongly associated with model predictions. These strongly associated attributes are identified as attribute-level explanations.
        To further quantify the contribution of each group, we apply a Shapley value–based explanation method (SHAP) to the surrogate, obtaining both global importance scores and per-patient attributions for each attribute group.

        \paragraph{Temporal segmentation.}
        To relate attribute-level explanations to collaboration dynamics, we examine how these attributes evolve throughout treatment. Using per‐snapshot node attributes, we construct temporal trajectories that capture changes in their magnitude over time (e.g., fluctuations in the involvement of \emph{Surgery} specialists). For each patient, we identify a surge point, defined as the snapshot at which the absolute change of a given attribute’s explanation value is maximal.
        % , subject to a minimum time threshold constraint to avoid trivial early cuts.
        Around this data-driven surge point, we apply a fixed-length temporal mask that partitions neighboring snapshots into three segments: a pre‐segment (five weeks), a transition segment centered at the surge point (three weeks), and a post‐segment (five weeks). This segmentation enables us to localize when an attribute’s influence on survival becomes most prominent by evaluating how the predictive performance varies across the three segments.
        
        For each attribute explanation (e.g., \emph{Surgery} or \emph{Oncology} involvement), the collaboration network snapshots are segmented using the corresponding surge point, and predictions are generated for all segments using the trained EvolveGCN. By comparing performance metrics such as F1 score across segments, we assess whether a given attribute truly drives the model toward better or worse survival predictions. For instance, attributes identified by the surrogate and SHAP analyses as beneficial are expected to yield increasingly higher predictive performance from the pre‐ through post‐segments, whereas attributes associated with poorer outcomes should show decreasing predicted survival probabilities.
        This combined surrogate and segmentation‐based framework provides a structured way to interpret TGNN predictions in terms of team collaboration and their evolution over the course of cancer care.

\section{Experiments}
\label{sec:exp}
    % how well the TGNN can predict patient survival given the patient-level collaboration network
    % We evaluate our framework on three disease-specific cohorts: breast cancer, lung/bronchus non–small cell, and colorectal cancer. For each cohort, we construct patient-level temporal collaboration networks and train a separate EvolveGCN-O model using the architecture and training procedure described in \autoref{sec:data} and \autoref{sec:methods}. 

    In this section, we demonstrate TGNN’s performance in predicting patient survival from EHR-mediated collaboration networks across different cancer types.
    % and how stable and interpretable the resulting patterns are across cancer cohorts. 
    % The constructed collaboration networks are split into training, validation, and test sets at the patient level to avoid information leakage. We use the same configuration across cohorts: binary classification with survival as the target, oversampling of the minority class where needed, and validation-based threshold tuning with F1 as the optimization metric. 
    For interpretability, following the procedure described in \autoref{sec:interpretMethod}, we train cohort-specific surrogate models to obtain attribute-level explanations. We further compute Shapley values \cite{lundberg2017unified} to validate and complement the explanations provided by the surrogate model. Together, these components enable us to identify which collaboration factors are most strongly associated with patient survival.

    \subsection{Survival Prediction Performance}
        % prediction performance
        % cross validation
        As described in \autoref{sec:predictMethod}, we measure the predictive performance by F1 scores with complementary metrics. \autoref{tab:tgnn_performance} reports F1-score, accuracy, precision, and recall on the held-out test sets.
        Across all three cohorts, the trained TGNN achieves non-trivial F1 scores and accuracies, indicating that survival outcomes can be meaningfully predicted from EHR-mediated collaboration patterns. 
        % Although balanced accuracy is lower in the Breast and Colorectal cancer cohorts—primarily due to the skewed distribution of alive versus deceased patients—this is expected, as class imbalance makes minority-class recall highly sensitive to even a small number of misclassifications. In contrast, the more comparable balanced accuracy in the Lung cancer cohort, together with consistently strong F1 scores across all cohorts, supports the conclusion that the model’s predictions are robust and maintain a reasonable balance between precision and recall (i.e., high F1). 
        Overall, the classifier effectively distinguishes networks belonging to alive and deceased patients. 
        % Balanced accuracy is reasonably close to overall accuracy, suggesting that the model is not simply defaulting to the majority class and that both survivors and non-survivors are being captured to some extent.

        % To assess robustness to the particular train/validation/test split, we also perform five-fold cross-validation within each cohort using the same training and threshold-tuning protocol. The fold-wise results (Table~\ref{tab:tgnn_cv}) show that mean F1 and accuracy are comparable to the single-split results, with moderate variance across folds that is consistent with the limited cohort sizes. This provides evidence that the observed performance is not an artifact of a single favorable data partition.

        % In later subsections, we further break down performance by temporal segment (pre, transition, post) using the dynamic segmentation scheme introduced in the Methodology, in order to understand where in the care trajectory the model’s predictions are most informative.
    
        \begin{table}[t]
        \centering
        \caption{Performance of TGNN for each cancer cohort.}
        \label{tab:tgnn_performance}
        \begin{tabular}{lcccc}
            \toprule
            Cohort & Accuracy & F1 & Precision & Recall \\
            \midrule
            Breast        &  0.867  &  0.927 & 0.884 & 0.974 \\
            Colorectal    &  0.767  &  0.863 & 0.815 & 0.917 \\
            Lung          &  0.821  &  0.848 & 0.875 & 0.824 \\
            \bottomrule
        \end{tabular}
    \end{table}

    \subsection{Survival Prediction Interpretation}
    \subsubsection{Attribute-based surrogate model performance. }
        % prediction performance
        % robustness evaluation
        % alignment evaluation
        
        To interpret what the TGNN has learned, we first train attribute-based surrogate models, as described in \autoref{sec:interpretMethod}, and evaluate how well their predictive behavior aligns with that of the TGNN.
        % For each cancer cohort, we aggregate node attributes into patient-level features that summarize the relative involvement of high-level collaboration groups (e.g., surgery, oncology, internal medicine, case management, primary care) and fit a logistic regression model on the same train/validation/test splits used for EvolveGCN-O.
        % The surrogate is optimized directly on the binary survival labels using cross-entropy loss. 
        We assess surrogate performance on the held-out test set using the same metrics applied to the TGNN. As shown in \autoref{tab:surrogate_performance}, the surrogate models achieve predictive performance that is closely aligned with—and only slightly below—that of the TGNN across all three cohorts. This suggests that a substantial portion of the survival signal captured by the TGNN can be represented using HCP and note attributes, supporting the use of surrogate models as interpretable proxies.

       \begin{table}[t]
            \centering
            \caption{Performance of the attribute-based surrogate model for each cancer cohort (test set).}
            \label{tab:surrogate_performance}
            \begin{tabular}{lcccc}
                \toprule
                Cohort & Accuracy & Precision &  Recall  & F1  \\
                \midrule
                Breast        & 0.962 & 0.773 & 1.000 & 0.872  \\
                Colorectal    & 1.000 & 1.000 & 1.000 & 1.000  \\
                Lung          & 0.932 & 0.919 & 0.919 & 0.919  \\
                \bottomrule
            \end{tabular}
        \end{table}

        % A summary of surrogate performance by cohort is provided in Table~\ref{tab:surrogate_performance}. 
        % (MOVED to robustness eval)Additionally, we train each cohort-specific surrogate with multiple random seeds and observe that the relative importance ranking of key groups (such as surgery, radiation, oncology, and internal medicine) remains unchanged across runs (see Evaluation section). This stability supports using the surrogate coefficients, together with SHAP-based importance scores, as robust estimates of how each group correlates with survival. In the following subsections, we combine the sign of these coefficients (positive versus negative correlation) with temporal segmentation to reason about how changes in group-specific involvement should manifest in pre, transition, and post segment performance.

    \subsubsection{Attribute-level explanation and temporal segmentation.}
        % using TGNN surrogate as a proxy (justify that all surrogate-based approaches adopt this framework)
        % We do so by using the surrogate as a proxy to identify influential collaboration groups and then relating these groups back to the temporal evolution of the collaboration network via our segmentation procedure.
        \paragraph{Attribute explanations.}
        Having established that the TGNN can predict survival with non-trivial accuracy and that an attribute-based surrogate models can recover much of this signal, we next investigate \textbf{what} the model has learned. Using the cohort-specific surrogate models, we examine the learned coefficients for each HCP and note attribute (i.e., the inputs to the surrogate) together with their corresponding Shapley values. We then compute per-patient contribution scores and aggregate them (via mean absolute value) to derive a global importance ranking of attributes.

        Across cohorts, a consistent set of influential attributes emerges. Higher involvement of Surgery, Oncology, and Internal Medicine specialists repeatedly ranks among the most important factors. In the breast and lung cohorts, Surgery exhibits one of the highest global importance scores with positive coefficients, indicating that increased surgical involvement is associated with better survival. Oncology involvement similarly appears as a key positive factor. In contrast, Internal Medicine involvement often receives negative coefficients across cohorts, suggesting an association with poorer outcomes.
        
        % (Moved to eval)Overall, these surrogate-based explanations suggest that the TGNN is leveraging clinically plausible features of team composition rather than spurious artifacts. In particular, the positive association between surgery involvement and survival in breast and lung cancer aligns with the idea that patients who reach timely, definitive local treatment are more likely to have early-stage, potentially curable disease.

        \paragraph{Segmenting network snapshots.}
        To connect attribute-level explanations (e.g., Surgery, Oncology, and Internal Medicine involvement) to collaboration dynamics, we construct temporal trajectories that capture how each attribute value evolves across network snapshots. As described in \autoref{sec:interpretMethod}, for each \emph{(patient, attribute)} pair, we identify a data-driven surge point—the snapshot at which the \emph{largest absolute} change between consecutive snapshots occurs. This surge point marks \textbf{when} a given important attribute begins to exert influence on the collaboration outcome. Additionally, around this surge point, we apply a fixed-length temporal mask that partitions the trajectory into three segments: a five-snapshot pre-segment, a three-snapshot transition segment, and a five-snapshot post-segment. This procedure is repeated independently for each attribute explanation and for each cancer cohort.
        
        % For each \emph{(cohort, group)}, we then record three quantities that we use in the subsequent analysis: (1) the sign of the surrogate coefficient (positive versus negative correlation with survival), (2) the direction of the change at the cut point (whether the group’s score increases or decreases after the cut), and (3) the segment-level predictive performance when EvolveGCN-O is evaluated on pre, transition, and post segments. These quantities allow us to relate changes in group-specific involvement to changes in the model’s predictive behavior along the treatment timeline.

        \paragraph{Segment performance: pre-segment, transition segment, post-segment}
        %to-be implement later once metrics has been finalized 
        % cumulative
        % within segment
        % pos corr: increasing performance
        % neg corr: decreasing performance

        Finally, we evaluate segment-level performance by running the trained TGNN on the pre-, transition-, and post-segments. In this analysis, we focus specifically on surge points derived from changes in the Surgery attribute. For each cohort, we generate predictions using only the snapshots that fall within a given segment and compute accuracy, precision, recall, and F1 score, on the held-out test patients.

        As shown in \autoref{tab:surgery_segments}, across all three cohorts, F1 scores in the transition segment are consistently higher than or comparable to those in the pre-segment. In the Breast and Lung cohorts, both F1 score and accuracy show substantial increases from the pre- to post-segment. In contrast, the Colorectal cohort exhibits relatively stable performance across segments, suggesting that changes in surgical involvement have a less pronounced effect for this group.

        % To further ground these findings in concrete examples, Figure~\ref{fig:network_snapshots} 
        % presents temporal snapshots of the EHR-mediated collaboration networks for two representative 
        % breast cancer patients—one survivor and one non-survivor—selected to be comparable in timeline 
        % length and number of surgical specialists involved in their care.
        \begin{figure*}[t]
            \centering
            \includegraphics[width=\linewidth]{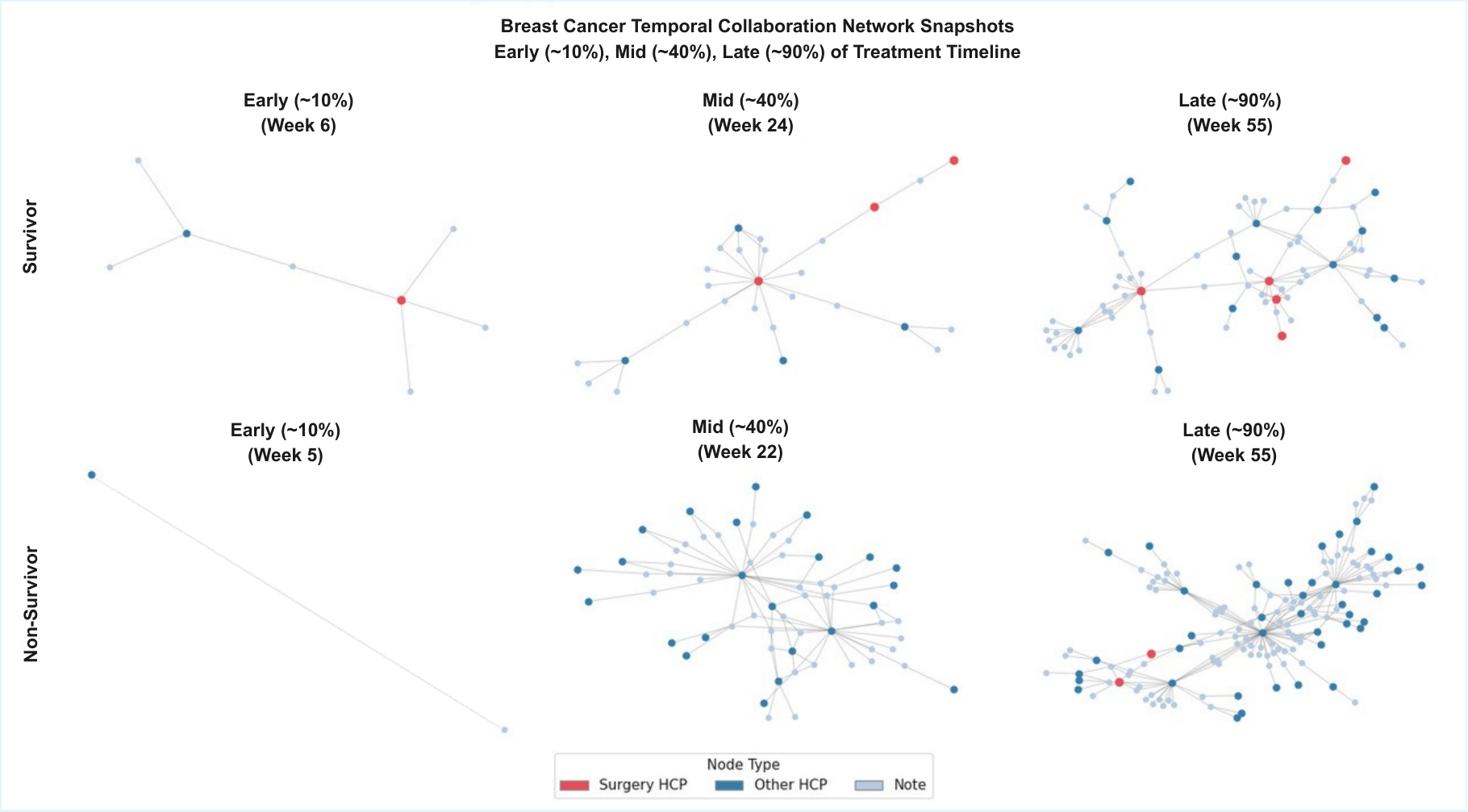}
            \caption{Temporal snapshots of EHR-mediated collaboration networks for a representative breast cancer survivor (top) and non-survivor (bottom) at early (10\%), mid (40\%), and late (90\%) of the treatment timeline. Each panel shows the largest connected component of the cumulative bipartite network at the corresponding snapshot. Nodes represent Surgery HCPs (red), other HCPs (dark blue), and clinical notes (light blue); edges connect HCPs to the notes they authored or reviewed. The survivor exhibits surgical specialist involvement from the early stage of treatment, whereas the non-survivor's network does not incorporate Surgery HCPs until late in the timeline, a pattern consistent with the positive association between surgical involvement and survival identified by our framework.}
            \label{fig:breast_network_snapshots}
        \end{figure*}
        
        To further ground these findings in concrete examples, \autoref{fig:breast_network_snapshots} presents temporal snapshots of the EHR-mediated collaboration networks for two representative breast cancer patients, one survivor and one non-survivor. These patients were selected based on three comparability criteria: similar treatment timeline lengths (62 and 55 weeks, respectively), a comparable number of surgical specialists present in their networks (5 and 2 nodes, respectively), and a contrasting timing of first surgical involvement (Week 5 versus Week 40). This selection ensures that the observed difference in outcomes is more likely attributable to \textit{when} surgical specialists were integrated into the care team, rather than differences in overall care complexity or team composition. Each sequence depicts the largest connected component of the cumulative bipartite network at early (10\%), mid (40\%), and late (90\%) of the treatment timeline. In the survivor's network, Surgery HCPs are integrated into the collaboration as early as Week 5 and are already active in the early-stage snapshot at Week 6, maintaining active connections throughout the course of treatment. In contrast, the non-survivor's network contains no surgical involvement until Week 40, with Surgery HCPs appearing only in the late stage. This contrast serves as a concrete illustration of the collaboration dynamic identified by our framework, namely that earlier integration of surgical specialists into the care team may be associated with improved survival outcomes. We note that this example is intended as an illustrative case rather than a definitive conclusion; broader empirical support is provided through the surrogate coefficients, SHAP rankings, and segment-level performance reported above.
      
      \begin{figure*}[t]
            \centering
            \includegraphics[width=\linewidth]{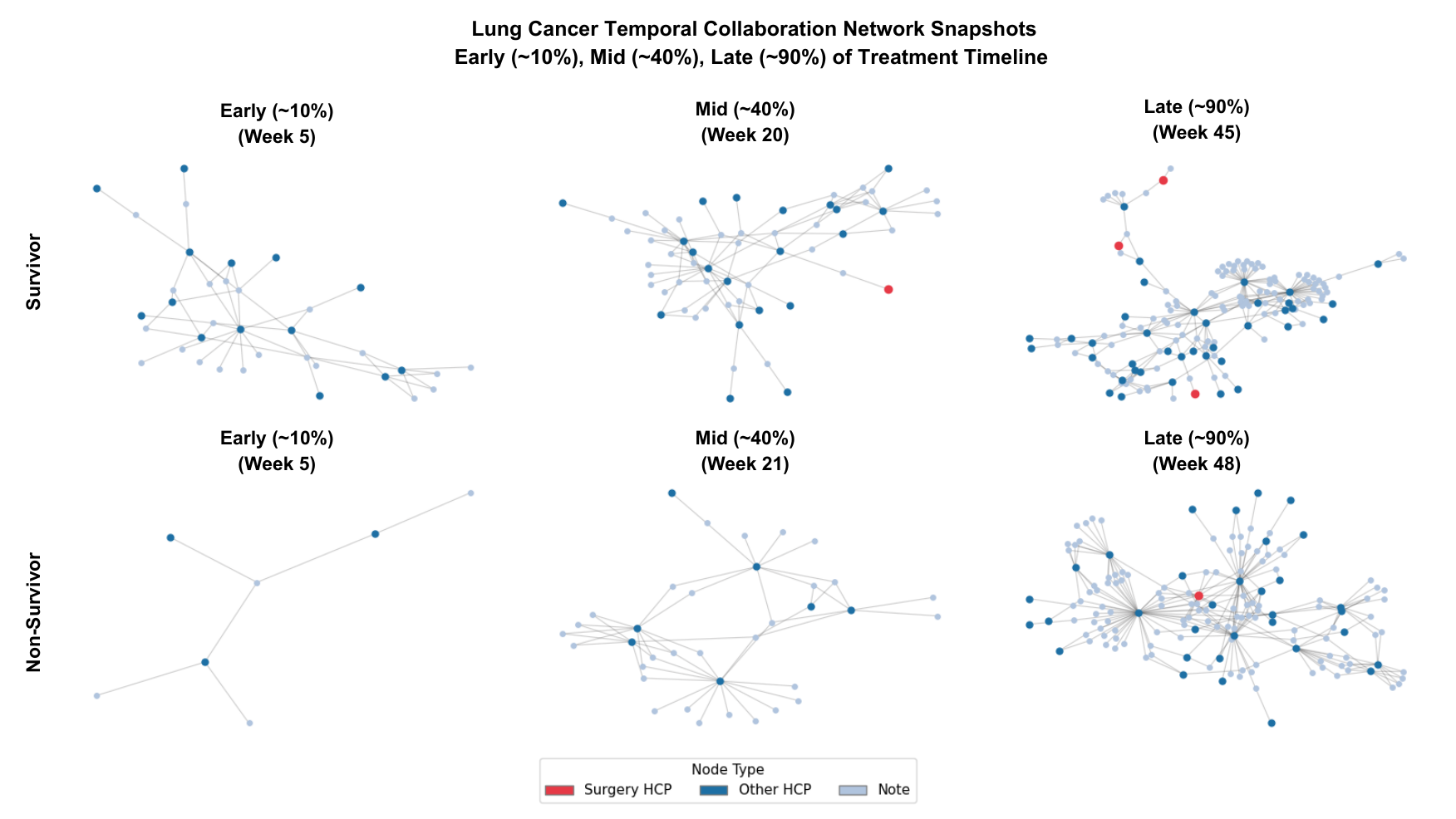}
                        \caption{Temporal snapshots of EHR-mediated collaboration networks for a representative lung cancer survivor (top) and non-survivor (bottom) at early (10\%), mid (40\%), and late (90\%) of the treatment timeline. Nodes represent Surgery HCPs (red), other HCPs (dark blue), and clinical notes (light blue). The survivor exhibits surgical specialist involvement from the early stage of treatment, whereas the non-survivor's network does not incorporate Surgery HCPs until late in the timeline, consistent with the positive association between surgical involvement and survival identified in the lung cohort.}
            \label{fig:lung_network_snapshots}
        \end{figure*}
       
        A similar pattern is observed in the lung cancer cohort. \autoref{fig:lung_network_snapshots} presents temporal snapshots for two representative lung cancer patients, one survivor and one non-survivor, selected using the same comparability criteria: similar treatment timeline lengths (51 and 54 weeks, respectively), a comparable number of surgical specialists present in their networks (3 and 1 nodes, respectively), and a contrasting timing of first surgical involvement (Week 7 versus Week 31). In the survivor's network, Surgery HCPs are integrated into the collaboration as early as Week 7 and remain active throughout treatment, as evidenced by their presence across all three snapshots. The non-survivor's network, by contrast, shows no surgical involvement until Week 31, after the mid-stage snapshot (Week 21), with Surgery HCPs appearing only in the late-stage snapshot (Week 48). This observation is consistent with the positive association between surgical involvement and survival identified in the lung cohort, as supported by the surrogate coefficients and SHAP rankings reported in \autoref{tab:surgery_importance}.

        Returning to the cohort-level analysis, these trends indicate that when the temporal window is centered around major increases in surgical involvement, the TGNN's predictions more strongly favor survival in the transition and post-segments than in the pre-segment. In other words, increased participation of Surgery specialists appears to shift collaborative patterns toward those associated with improved outcomes for Breast and Lung cancer patients, consistent with the positive coefficients assigned to this attribute by our surrogate model.
        %(MOVED to domain eval) This is consistent with our interpretation of surgery as a key phase in the care trajectory for early-stage, potentially curable cancers, where collaboration patterns around the surgical period carry strong prognostic information.

         \begin{table}[t]
                \centering
                \caption{Segment-level performance of TGNN on surgery-guided segments (test set).}
                \label{tab:surgery_segments}
                \begin{tabular}{lccccc}
                    \toprule
                    Cohort & Segment & Accuracy & Precision & Recall & F1 \\
                    \midrule
                    Breast       & Pre         & 0.769 & 0.906 & 0.829 & 0.866 \\
                                 & Transition  & 0.923 & 0.921 & 1.000 & 0.959 \\
                                 & Post        & 0.897 & 0.897 & 1.000 & 0.946 \\
                    \midrule
                    Colorectal   & Pre         & 0.786 & 0.786 & 1.000 & 0.880 \\
                                 & Transition  & 0.786 & 0.786 & 1.000 & 0.880 \\
                                 & Post        & 0.778 & 0.778 & 1.000 & 0.875 \\
                    \midrule
                    Lung         & Pre         & 0.576 & 0.640 & 0.762 & 0.696 \\
                                 & Transition  & 0.636 & 0.645 & 0.952 & 0.769 \\
                                 & Post        & 0.656 & 0.778 & 0.667 & 0.718 \\
                    \bottomrule
                \end{tabular}
                \end{table}

         % \begin{table}[t]
         %        \centering
         %        \caption{Segment-level performance of TGNN on surgery-guided segments (test set).}
         %        \label{tab:surgery_segments}
         %        \begin{tabular}{lcccccc}
         %            \toprule
         %            Cohort & Segment & Accuracy & Precision & Recall & F1 & Balanced accuracy \\
         %            \midrule
         %            Breast       & Pre         & 0.769 & 0.906 & 0.829 & 0.866 & 0.539 \\
         %                         & Transition  & 0.923 & 0.921 & 1.000 & 0.959 & 0.625 \\
         %                         & Post        & 0.897 & 0.897 & 1.000 & 0.946 & 0.500 \\
         %            \midrule
         %            Colorectal   & Pre         & 0.786 & 0.786 & 1.000 & 0.880 & 0.500 \\
         %                         & Transition  & 0.786 & 0.786 & 1.000 & 0.880 & 0.500 \\
         %                         & Post        & 0.778 & 0.778 & 1.000 & 0.875 & 0.500 \\
         %            \midrule
         %            Lung         & Pre         & 0.576 & 0.640 & 0.762 & 0.696 & 0.506 \\
         %                         & Transition  & 0.636 & 0.645 & 0.952 & 0.769 & 0.518 \\
         %                         & Post        & 0.656 & 0.778 & 0.667 & 0.718 & 0.652 \\
         %            \bottomrule
         %        \end{tabular}
         %        \end{table} 

\section{Evaluation}
    % from supporting literature
    % our results are the first data-diven empirical evidence for these hypotheses
    \label{sec:evaluation}

    In this section, we evaluation (1) the robustness of our surrogate models and (2) the validity if the identified important dynamic collaboration traits.
    
    % We now evaluate whether the patterns learned by our framework are consistent with prior clinical knowledge about surgery and survival in early-stage cancer. We focus on surgery because it repeatedly emerges as a stable, positive factor across cohorts in our surrogate, Shapley (SHAP) values, and segment-level analyses.
    
    \subsection{Robustness Evaluation}
    To evaluate the robustness of our surrogate models, we re-train each surrogate model using ten random seed. Across the ten runs, the rank of attributes per cohort derived from the learned weights are the same, indicating their relative importance for a cohort is robust to training dynamics and the consistent rank is stable and not affected by the stochastic nature of machine learning approaches. \autoref{tab:surgery_importance} reports the importance scores (i.e., learned weights and Shapley values) and the respective rank of the attribute ``Surgery'' among all attributes averaged across ten runs, indicating its consistent and high influence on the patient survival. 
    % First, we check how surgery behaves in the cohort-specific logistic regression surrogate. Across ten random seeds per cohort, the learned coefficients for surgery and other treatment groups are numerically identical, indicating that the surrogate solution is stable to stochastic variation in the train/validation split. For example, in the breast cohort the coefficient for surgery is consistently $0.0043$, while radiation is $0.0034$ in all runs. In the lung (NSCLC) cohort, surgery has a positive coefficient of $0.0022$ and radiation has $0.0030$ across all seeds. In colorectal cancer, surgery remains positive ($0.00075$), with radiation and oncology having slightly larger positive weights (both around $0.0013$). 

    % the cohort-level surgery coefficient (averaged over seeds, though in practice identical across runs) together with its global SHAP rank and mean absolute SHAP value. In breast and lung cancer, surgery has the highest global importance among clinician groups (SHAP rank~1 in both cohorts) with large mean absolute SHAP values ($0.68$ for breast, $0.46$ for lung), and in colorectal cancer it still appears as a non-negligible positive factor (rank~5, mean absolute SHAP $\approx 0.11$). Together, the positive coefficients and high SHAP importance suggest that greater surgical involvement pushes the surrogate’s predictions toward survival rather than death.
    
    \begin{table}[t]
        \centering
        \caption{Statistics of the attribute ``Surgery'': logistic regression (LR) coefficient, LR coefficient rank among all attributes, mean absolute SHAP value, and SHAP value rank among all attributes.}
        \label{tab:surgery_importance}
        \begin{tabular}{lcccc}
            \toprule
            Cohort & LR weight & LR rank & Mean $|\text{SHAP}|$ & SHAP rank \\
            \midrule
            Breast        & 0.0043 & 1 & 0.68 & 1 \\
            Lung          & 0.0022 & 2 & 0.46 & 1 \\
            Colorectal    & 0.0007 & 3 & 0.11 & 5 \\
            \bottomrule
        \end{tabular}
    \end{table}

    \subsection{Insight Validity}
    % \subsection{Consistency with clinical evidence on early-stage cancer}
    We compare our data-driven findings with established clinical literature. For Lung cancer, recent evidence~\cite{sts_lung_resection_2025} shows that anatomic lung resections, such as lobectomy and segmentectomy, yield 5-year overall survival rates of approximately 70–72\%. For Breast cancer, surgery remains the primary treatment for stages 1 through 3, often in combination with other modalities; although survival rates vary slightly across surgical techniques~\cite{fisher2015survival}, surgical intervention consistently represents an effective component of care. These findings support the results generated by our framework: positive surgery coefficients in the surrogate model, high global SHAP importance for breast and lung cohorts, and improved segment-level performance around increases in surgical involvement.

    Taken together, our results align with well-established clinical understanding that timely, surgical treatment is central to curative treatment for Breast cancer and resectable Lung cancer. Moreover, our framework has the potential to extend these insights by indicating \textbf{how early} surgical specialists should be involved to achieve optimal effectiveness, based on real-world patient data. Our method provides a data-driven pipeline for future advances in clinical decision-making and intervention planning.
    
    % Our framework does not use these survival rates directly, but it independently predicted that increased surgery involvement is associated with better survival in both breast and lung cohorts, and that model performance is strong in segments centered around surgery-related changes. The positive surgery coefficients, high global SHAP importance in breast and lung, and favorable segment-level metrics aligns with established clinical understanding: timely, definitive surgery is a key component of curative treatment for early-stage breast and early-stage resectable lung cancer. In this sense, our results provide similar data-driven evidence that EHR-mediated collaboration patterns around surgery lead to a patient prognosis, rather than reflecting spurious predictions. 

\section{Discussion and Future Works}
In \autoref{sec:method} and \autoref{sec:exp}, we introduced several experimental choices, including our selection of the TGNN model and the temporal window lengths used for snapshot segmentation. We emphasize that these decisions are not intended as prescriptive guidelines but rather as one viable experimental setup that enables the analysis of collaboration dynamics and the extraction of meaningful, interpretable insights. In this section, we discuss design choices that warrant further exploration and outline avenues for future research.

Many existing TGNN models combine graph neural network (GNN) and recurrent neural network (RNN) components sequentially—typically encoding structure with a GNN and then capturing dynamics with an RNN \cite{feng2025comprehensive}. EvolveGCN, however, integrates these mechanisms more tightly: an RNN updates the internal GNN parameters at each time step. This design improves performance but makes the model behavior more opaque, as neither RNN attention weights nor GNN parameters alone provide a reliable proxy for snapshot-level importance. To address this challenge, we developed a model-agnostic interpretation procedure that does not rely on architectural interpretability and can be applied to a broad class of TGNN models. Future work includes directly analyzing the internal RNN and GNN weights to derive comprehensible network dynamics associating with the EvolveGCN model behavior.

Another avenue for exploration is systematically varying the temporal window length used in snapshot segmentation. Although our current choices yield meaningful insights, a more extensive evaluation is needed to understand how window size influences interpretability and model performance.
Finally, in this work we focus on attributes that consistently receive high surrogate-model rankings across cancer types, using them to study how their temporal patterns relate to survival. A future direction is to enable interactive selection of these attributes through an expert-facing interface, allowing clinicians to explore ranked attributes directly and test their own hypotheses.
\section{Conclusion}
This paper introduces a data-driven approach for modeling and interpreting how clinical teams collaborate over the course of cancer care. By leveraging EHR traces, we construct temporal collaboration networks for individual patients and apply a temporal graph neural network to learn survival-associated patterns within these evolving structures. We further propose an interpretation procedure that links model decisions to specific teamwork behaviors and identifies when they become influential.
Our results show that the proposed framework achieves stable and generalizable predictive performance while yielding insights that align with existing medical literature. Using real-world data, our analysis provides empirical evidence that earlier involvement of surgical specialists in the care trajectory may improve outcomes, and our temporal interpretation further pinpoints how early this engagement should occur, paving the way for future research on implementing such practices in real-world care.
In summary, this work demonstrates how machine learning and digital collaboration data can be used not only to predict outcomes but also to guide actionable, time-sensitive interventions in team-based cancer care. The framework offers a generalizable pathway for examining collaborative dynamics in other high-stakes clinical domains where coordination among HCPs is central to patient outcomes.

\begin{acks}
    This work was supported by grant R01CA273058 and R01CA270454 from the National Cancer Institute. Contents of this manuscript are solely the responsibility of the authors and do not represent the official view of the National Cancer Institute.
\end{acks}

\bibliographystyle{ACM-Reference-Format}
\bibliography{00_ref}

\end{document}